\documentclass[aps,prc,preprint,superscriptaddress,showpacs,amsmath,floatfix]{revtex4}

\usepackage{graphicx}
\usepackage{longtable}

\begin{document}


\title{
Polarization transfer in the $^{16}$O$(p,p')$ reaction at forward angles
and structure of the spin-dipole resonances}

\author{T.~Kawabata}
\email[]{takahiro@rcnp.osaka-u.ac.jp}
\altaffiliation[Present address:]{
Research Center for Nuclear Physics, 
Osaka University, Ibaraki, Osaka 567-0047, Japan}
\affiliation{Department of Physics, Kyoto University,
Kyoto 606-8502, Japan}
\author{H.~Akimune}
\affiliation{Department of Physics, Konan University,
Kobe, Hyogo 658-8501, Japan}
\author{G.P.A.~Berg}
\affiliation{Research Center for Nuclear Physics, 
Osaka University, Ibaraki, Osaka 567-0047, Japan}
\author{H.~Fujimura}
\affiliation{Research Center for Nuclear Physics, 
Osaka University, Ibaraki, Osaka 567-0047, Japan}
\author{H.~Fujita}
\affiliation{Department of Physics, Osaka University,
Toyonaka, Osaka 560-0043, Japan}
\author{Y.~Fujita}
\affiliation{Department of Physics, Osaka University,
Toyonaka, Osaka 560-0043, Japan}
\author{M.~Fujiwara}
\affiliation{Research Center for Nuclear Physics, 
Osaka University, Ibaraki, Osaka 567-0047, Japan}
\affiliation{Advanced Science Research Center,
Japan Atomic Energy Research Institute,
Tokai, Ibaraki 319-1195, Japan}
\author{K.~Hara}
\affiliation{Research Center for Nuclear Physics, 
Osaka University, Ibaraki, Osaka 567-0047, Japan}
\author{K.~Hatanaka}
\affiliation{Research Center for Nuclear Physics, 
Osaka University, Ibaraki, Osaka 567-0047, Japan}
\author{K.~Hosono}
\affiliation{Department of Engineering, Himeji Institute of Technology,
Hyogo 678-1297, Japan}
\author{T.~Ishikawa}
\affiliation{Department of Physics, Kyoto University,
Kyoto 606-8502, Japan}
\author{M.~Itoh}
\affiliation{Department of Physics, Kyoto University,
Kyoto 606-8502, Japan}
\author{J.~Kamiya}
\affiliation{Research Center for Nuclear Physics, 
Osaka University, Ibaraki, Osaka 567-0047, Japan}
\author{M.~Nakamura}
\affiliation{Department of Physics, Kyoto University,
Kyoto 606-8502, Japan}
\author{T.~Noro}
\altaffiliation[Present address:]{
Department of Physics, Kyushu University, Fukuoka 812-8581, Japan}
\affiliation{Research Center for Nuclear Physics, 
Osaka University, Ibaraki, Osaka 567-0047, Japan}
\author{E.~Obayashi}
\affiliation{Research Center for Nuclear Physics, 
Osaka University, Ibaraki, Osaka 567-0047, Japan}
\author{H.~Sakaguchi}
\affiliation{Department of Physics, Kyoto University,
Kyoto 606-8502, Japan}
\author{Y.~Shimbara}
\affiliation{Department of Physics, Osaka University,
Toyonaka, Osaka 560-0043, Japan}
\author{H.~Takeda}
\affiliation{Department of Physics, Kyoto University,
Kyoto 606-8502, Japan}
\author{T.~Taki}
\altaffiliation[Present address:]{
Asaka Technology Development Center, 
Fuji Photo Film Co., Ltd., Asaka, Saitama 351-8585, Japan}
\affiliation{Department of Physics, Kyoto University,
Kyoto 606-8502, Japan}
\author{A.~Tamii}
\affiliation{Department of Physics, University of Tokyo, Hongo,
Tokyo 113-0033, Japan}
\author{H.~Toyokawa}
\affiliation{Japan Synchrotron Radiation Research Institute, 
Hyogo 679-5198, Japan}
\author{M.~Uchida}
\affiliation{Department of Physics, Kyoto University,
Kyoto 606-8502, Japan}
\author{H.~Ueno}
\altaffiliation[Present address:]{
RIKEN (The Institute for Physical and Chemical Research),
Wako, Saitama 351-0198, Japan}
\affiliation{Department of Physics, Osaka University,
Toyonaka, Osaka 560-0043, Japan}
\author{T.~Wakasa}
\affiliation{Research Center for Nuclear Physics, 
Osaka University, Ibaraki, Osaka 567-0047, Japan}
\author{K.~Yamasaki}
\affiliation{Department of Physics, Konan University,
Kobe, Hyogo 658-8501, Japan}
\author{Y.~Yasuda}
\affiliation{Department of Physics, Kyoto University,
Kyoto 606-8502, Japan}
\author{H.P.~Yoshida}
\affiliation{Research Center for Nuclear Physics, 
Osaka University, Ibaraki, Osaka 567-0047, Japan}
\author{M.~Yosoi}
\affiliation{Department of Physics, Kyoto University,
Kyoto 606-8502, Japan}

\date{\today}

\begin{abstract}
Cross sections and polarization transfer observables in the
$^{16}$O$(p,p')$ reactions at 392 MeV were measured
at several angles between $\theta_{lab}=$ 0$^\circ$ and
14$^\circ$. The non-spin-flip (${\Delta}S=0$) and spin-flip
(${\Delta}S=1$) strengths in transitions to several discrete states and
broad resonances in $^{16}$O were extracted using a model-independent
method. The giant resonances in the energy region of
$E_x=$19$-$27 MeV were found to be predominantly excited by
${\Delta}L=1$ transitions. The strength distribution of spin-dipole
transitions with ${\Delta}S=1$ and ${\Delta}L=1$ were deduced. The
obtained distribution was compared with a recent shell model
calculation. Experimental results are reasonably explained by
distorted-wave impulse approximation calculations with the shell model
wave functions.
\end{abstract}

\pacs{
24.30.Cz;
24.70.+s;
25.40.Ep;
27.20.+n;
}

\maketitle


\section{Introduction}
Spin-isospin excitation modes in nuclei have been studied intensively,
not only because they are of interest in nuclear structure, but also
because the relevant operators mediate $\beta$-decay and neutrino
capture processes. The cross sections of hadronic reactions provide a
good measure for the weak interaction response, which is a key
ingredient in studies of nucleosynthesis. Gamow-Teller resonances
(GTR; $\Delta T=1$, $\Delta S=1$, $\Delta L=0$) mediated by the
$\vec{\sigma}\vec{\tau}$ operator have been systematically
investigated by charge exchange reactions like $(p,n)$ and
$(^3$He,$t)$ reactions with a selectivity for spin-flip transitions at
intermediate energies \cite{ALFO98,MAMO96}. On the other hand,
spin-dipole resonances (SDR; $\Delta T=1$, $\Delta S=1$, $\Delta L=1$)
mediated by the $\vec{\sigma}\vec{\tau}rY_1$ operator have not been
studied in any detail although the excitations have recently received
attention from the view point of detection of supernova neutrinos
\cite{KOLB01,FULL99,LANG96}. The detailed structure of the SDR remains
unclear with respect to the three different spin states of
$J^\pi=2^-$, $1^-$, and $0^-$. 

Transitions to the $1^-$ states can be
induced by a probe with spin through the spin-flip and
non-spin-flip processes with the $\vec{\sigma}\vec{\tau}rY_1$ and
$\vec{\tau}rY_1$ operators. The $\vec{\tau}rY_1$ operator mediates the
isovector giant dipole resonance (IVGDR; $\Delta T=1$, $\Delta S=0$,
$\Delta L=1$), which has the spin-parity of $J^\pi=1^-$, the same as
the SDR. Theoretically, the SDR and IVGDR are observed together
in $(p,p')$ reactions because they have the same $J^\pi=1^{-}$ and are 
located in the same excitation energy region. The problem concerning
the coexistence of the SDR and IVGDR was discussed in
Ref.~\cite{SAGA85}.

The SDR in the A=12 system has been relatively well studied in the
past. Cross sections of the SDR in the $^{12}$C$(p,n)$ reaction were
measured at various bombarding energies and the strength distributions
were discussed and compared with shell model calculations in view of the
analog relation in N=Z nuclei \cite{GAAR84}. The experimental analysis 
of the data has led to the conclusion that broad structures at $E_x=$
4.2 and 7.2 MeV in $^{12}$N consist of mainly $2^-$ and $1^-$,
respectively. Recent measurements of $(p,n)$ and $(n,p)$ reactions
supported this conclusion \cite{YANG93}. An analysis of the tensor
analyzing powers measured in the $^{12}$C$(d,^2$He) reaction resulted
in the contradictory conclusion that the $2^-$ transition dominates
around $E_x=7.5$ MeV in $^{12}$B, which corresponds to the excitation
region around $E_x=7.2$ MeV in $^{12}$N, and the contribution from the
$1^-$ transition is small \cite{OKAM95}.  This ``{\it missing
spin-flip $1^-$}'' result was supported by a shell model study, which
predicted that the tensor correlation and the mixing of 2p-2h
configurations push the spin-flip $1^-$ strength to higher excitation
energies and thus quench the strength around the giant resonance
region \cite{SUZU98}. However, more recent measurements of
neutron-decay in the SDR region excited via the $(d,^2$He) reaction
supported the $(p,n)$ and $(n,p)$ results again \cite{INOM98}. Thus,
this problem concerning the spin-parity assignment for the SDR in the
A=12 system remains controversial.

The $^{16}$O nucleus consists of eight protons and eight neutrons in
the $1s_{1/2}$, $1p_{3/2}$, and $1p_{1/2}$ shell orbitals in a simple
independent particle model. Since the SDR excitation in
$p$-shell nuclei is described as a coherent sum of 1p-1h transitions
from the $p$- to the $sd$-shell orbitals, the SDR excitations in
$^{16}$O are expected to be stronger than those in $^{12}$C. Djalali
{\it et al.} identified several $2^-$ and $1^-$ states at
$E_x=$19$-$27 MeV in $^{16}$O by comparing a $(p,p')$ spectrum at
$E_{p}=201$ MeV with a $(\gamma,n)$ spectrum \cite{DJAL87}. They
pointed out that the gross structures of the $1^-$ resonances observed
in $(p,p')$ and $(\gamma,n)$ reactions are similar. This suggests that
the IVGDR, which is excited through the Coulomb interaction, is
dominant in the proton inelastic scattering at intermediate energies,
especially at forward angles. Therefore, spin-flip $1^-$ states were
not identified in the $(p,p')$ measurement of Ref.~\cite{DJAL87}. 

It is well known that the analog states of N=Z nuclei can be easily
observed in charge exchange reactions. 
Fazely {\it et al.} identified two strong $2^-$ states at $E_x=0.4$
and 7.6 MeV, and two broad $1^-$ states at 9.4 and 11.5 MeV in
$^{16}$F by measuring angular distributions of the $^{16}$O$(p,n)$
cross sections \cite{FAZE82}. In addition, Hicks {\it et al.}
\cite{HICK91} and Mercer {\it et al.} \cite{MERC94} reported that a
sizable amount of the ${\Delta}L=1$ strength in $^{16}$F and $^{16}$N
was observed in the $^{16}$O$(p,n)$ and $^{16}$O$(n,p)$ reactions
using a multipole decomposition method. In their multipole
decomposition analyses, spin-flip and non-spin-flip contributions of
the ${\Delta}L=1$ strength were not distinguished. Since the $(p,n)$
and $(n,p)$ reactions at intermediate energies dominantly carry the
spin-flip strength at forward angles, the ${\Delta}L=1$ strength in
the $^{16}$O$(p,n)$ and $^{16}$O$(n,p)$ reactions should include the
${\Delta}S=1$ (SDR) components. Watson {\it et al.} performed a high
resolution measurement of the $^{16}$O$(p,n)$ reaction at $E_{p}=135$
MeV and identified 2$^-$, 1$^-$, and 1$^+$ transition strengths by
measuring cross sections and vertical polarization transfer
observables (PT-observables) \cite{WATS94}. For further clarification
of the spin nature of the 1$^-$ transitions, it is, however, necessary
to measure horizontal PT-observables, too.

There is another important aspect in studying the level structure of
the excited states in $^{16}$O. The nucleus $^{16}$O is now considered
as a possible neutrino detector to investigate the explosion mechanism
of supernovae \cite{LANG96,KOLB92,EJIR00,SUZU00a,SIIS01}. High energy
neutrinos with an average energy of $\approx$25 MeV are emitted from
the heat bath in a supernova. When a supernova collapses near our
galaxy, it is expected that a sizable flux of neutrinos from the
neutronization process and the subsequent thermal emission process
appears on earth during a sub-second period. Such supernova neutrinos
can be detected by measuring deexcitations of the excited states in
$^{16}$O, $^{16}$F, and $^{16}$N in large neutrino detectors, {\it
e.g.} Superkamiokande and SNO with a huge number of $^{16}$O
nuclei. The neutrinos predominantly excite $0^-$, $1^-$, and $2^-$
states via neutral and charged current reactions \cite{LANG96}. Many
calculations were performed to estimate the cross sections of neutrino
induced reactions on $^{16}$O \cite{KOLB92, SUZU00a, SIIS01}. To
confirm these calculations, measurements of the strength distributions
of $0^-$, $1^-$, and $2^-$ states in $^{16}$O are important not only
for nuclear physics but also for the astrophysical application.

Recently, PT-observables for $(p,p')$ reactions were successfully
measured at 0$^\circ$ and were found to be a useful spectroscopic tool
to study nuclear structure \cite{TAMI99,ISHI01}, because the total
spin transfer $\Sigma\equiv[3-(D_{SS}+D_{NN}+D_{LL})]/4$, provides a
clear mean to clarify spin-flip or non-spin-flip transitions
\cite{SAKA98,SUZU00b}. Thus, measurements of PT-observables at forward
angles enable us to extract spin-flip transitions. In this report, we
will present information on the structure of the SDR in
$^{16}$O, which is obtained from measurements of PT-observables in
proton inelastic scattering at very forward angles including
0$^\circ$. 

\section{Experimental Procedure}

The experiment was performed at the Research Center for Nuclear
Physics (RCNP), Osaka University by using a 392 MeV polarized proton
beam accelerated by the coupled cyclotrons. The proton beam from the
polarized ion source \cite{HATA97} was accelerated to a kinetic energy
of 64.2 MeV
by the $K=120$ MeV AVF (Azimuthally Varying Field) cyclotron, and was
further boosted to 392 MeV by the $K$ = 400 MeV Ring cyclotron. The
proton beam extracted from the Ring cyclotron was achromatically
transported to the target. The beam intensity on target was in the
range of 1$-$10 nA. Scattered protons were momentum analyzed by
the high-resolution spectrometer
Grand Raiden \cite{MAMO99}. Trajectories of the scattered protons were
determined using a focal-plane detector system consisting of two
multi-wire drift chambers and plastic scintillation detectors. Cross
sections were obtained at seven scattering angles between
$\theta_{lab}=0^\circ$
and 14$^\circ$. An energy resolution of 80$-$150 keV full width
at half maximum (FWHM) was obtained, which was dominated by the energy 
spread of the cyclotron beam. In the measurements at scattering angles
between 6$^\circ$ and 14$^\circ$, the proton beam was stopped in a
Faraday cup in the scattering chamber. Since this Faraday cup reduced
the acceptance of the spectrometer Grand Raiden at angles between
2.5$^\circ$ and 4$^\circ$ and unacceptable background from edge
scattering was produced, a different Faraday cup was used between the
first quadrupole (Q1) and the sextupole (SX) magnets of Grand Raiden
\cite{MAMO99} at these angles. In the measurement at $0^\circ$, the
proton beam passed through Grand Raiden and was stopped in another
Faraday cup located 12 m downstream of the focal plane \cite{YOSO01}. 

\subsection{H$_2$O ice target}
\begin{figure}
\includegraphics[scale=0.44]{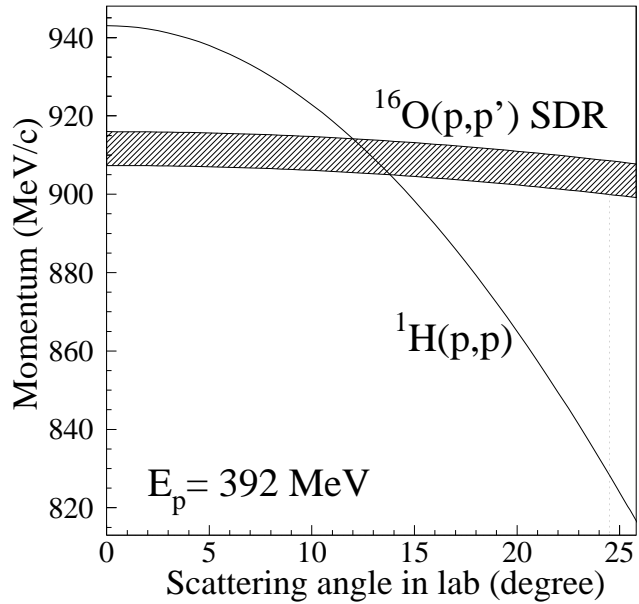}
\caption{\label{kinema}
Momentum of protons scattered from $^1$H and $^{16}$O at $E_{p}=392$
MeV as a function of angles. The hatched region shows the momentum
region of protons exciting the SDR (19 MeV $\le E_x \le$ 25 MeV).
}
\end{figure}

Thin H$_2$O ice sheets with various thicknesses of 10$-$30
mg/cm$^2$ were used as oxygen targets. The ice target system is
described in detail in Ref.~\cite{BATA01}. Here, we briefly describe
the target preparation procedure. Self supporting ice sheets made of
pure water were mounted in the scattering chamber, which was kept
under vacuum lower than 10$^{-3}$ Pa without any window-foil. The ice
targets were cooled down to below 140 K by liquid nitrogen. The loss
of the ice target by sublimation is negligible at this temperature,
since vapor-pressure of H$_2$O is of the order of 10$^{-6}$ Pa at 140
K and decreases exponentially at lower temperature. This newly
developed ice target gave us the great advantage in obtaining clean
$(p,p')$ spectra since the background events from hydrogen
contaminations in the target are out of the interested momentum region
at most of the measurement angles due to the large difference of
kinematic effects between oxygen and hydrogen. Fig.~\ref{kinema}
shows the kinematics of protons scattered from $^1$H and $^{16}$O as
a function of angles. In measurements for the SDR region
($E_x=$19$-$25 MeV) denoted by the hatched region in
Fig.~\ref{kinema}, the $^{16}$O$(p,p')$ events are obscured by the
$^1$H(p,p) events at $\theta_{lab}=$12$^\circ-$14$^\circ$, but are not
disturbed at the other angles.

In order to monitor the target thickness during the measurement,
elastic scattering events from hydrogen were measured by the
spectrometer LAS \cite{MATS95} placed at
$\theta_{lab}=59.5^\circ$. The target thickness
was calibrated by using the cross section for proton-proton elastic
scattering at $\theta_{lab} = 59.5^\circ$ calculated with the program
SAID \cite{ARND97}. The target thickness was stable within the
measurement uncertainty of $\pm2.5$\% during the experiment.

\subsection{Polarization transfer measurements}

A proton beam from the AVF cyclotron was vertically polarized. Two
super-conducting solenoids between the two cyclotrons were used for
the purpose of rotating the polarization axis of protons from the
vertical to the horizontal direction in measurements with a
horizontally polarized beam. The two solenoids were located
upstream and downstream of bending magnets with a total bending angle
of 45$^\circ$. This configuration enables us to obtain horizontally
polarized beams with two polarization axes approximately perpendicular
to each other. The beam polarization was 0.6$-$0.8, which was
monitored with an accuracy of $\pm0.02$ by two sets of beam-line
polarimeters after the Ring cyclotron using a polystyrene analyzer
target. The direction of the beam polarization was reversed every
second to eliminate instrumental asymmetries.

The polarization of protons scattered from the ice target were
measured at laboratory angles of 0$^\circ$, 4$^\circ$, and 8$^\circ$
by a focal plane polarimeter (FPP) after momentum analysis in the
spectrometer Grand
Raiden. The FPP consisted of an analyzer target of a 12 cm-thick
carbon slab, four multi-wire proportional chambers, and scintillator
hodoscopes \cite{YOSO95}.  The effective analyzing power $A_y^{eff}$ of
the FPP is given by
\begin{equation}
A_y^{eff}=
\frac{\int
\sigma^{inc}(\theta){A_y}^{inc}(\theta)\cos\phi d\Omega}
{\int\sigma^{inc}(\theta)d\Omega},
\label{eq:effay}
\end{equation}
where $\sigma^{inc}(\theta)$ and ${A_y}^{inc}(\theta)$ are
the cross section and analyzing power for inclusive proton scattering
from elastic, inelastic, and quasi-free processes in the analyzer
target of the FPP. The angular integrations in Eq.~(\ref{eq:effay})
are performed over the polar angle of $5^\circ\le\theta\le20^\circ$
and the azimuthal angle of $|\phi|\le66.8^\circ$ for the scattering
angles. The inclusive cross section $\sigma^{inc}(\theta)$ was
measured in this experiment. We used the analyzing power
${A_y}^{inc}(\theta)$ given in Ref.~\cite{MCNA85}, which is
parameterized as a function of the proton energy and scattering angle.

The PT-observables $(D_{I'J})$ are defined by the following relation
\cite{HOSH86,OHLS72},
\begin{widetext}
\begin{equation}
\left(
\begin{array}{c}{p'}_{S'}\\{p'}_{N'}\\{p'}_{L'}\\\end{array}\right)
=
\frac{1}{1+p_NA_N}
\left[
\left(\begin{array}{c}0\\P\\0\\ \end{array}\right)
+\left(\begin{array}{ccc}
D_{S'S}&        0       &D_{S'L}\\
0&              D_{N'N} &0\\
D_{L'S}&        0       &D_{L'L}\\
\end{array}\right)
\left(\begin{array}{c}p_{S}\\p_{N}\\p_{L}\\\end{array}\right)
\right].
\label{eq:ptdef}
\end{equation}
\end{widetext}
The symbols, ${p}_{I}$ and ${p'}_{I'}$ ($I=S$, $N$, $L$), denote the
polarizations of the incident and the scattered protons,
respectively. The coordinate system is chosen so that the $L$ axis is
along the beam direction, the $N$ axis is along the normal to the
horizontal plane, and the $S$ axis is chosen to form a right-handed
coordinate system (projectile helicity frame). Similarly, the $L'$
axis is oriented along the momentum of scattered protons, the $N'$
axis is the same as the $N$ axis, and the $S'$ axis forms a
right-handed coordinate system (outgoing particle helicity frame). The
symbols, $A_N$ and $P$, are analyzing power and vector polarization,
respectively. The off diagonal elements of PT-observables $D_{I'J}$
between the horizontal and vertical axes vanish due to parity
conservation. 

The proton spin precesses around the vertical axis of the
spectrometer. The spin precession angle $\chi$ with respect
to the momentum direction of the proton is described by
$\chi=\gamma(\frac{g}{2}-1)\alpha$ in the moving frame,
where $\gamma$ is a Lorentz factor defined by
$\gamma=\frac{m_pc^2+E_p}{m_pc^2}$, $g$ is the spin g-factor of 
the proton, which is related to the proton magnetic moment by
$\mu_p=\frac{1}{2}g\mu_N$ ($\mu_N$ = nuclear magneton), and $\alpha$
is the bending angle of the spectrometer. The vertical ($p'_{N''}$)
and horizontal ($p'_{S''}$) components of the polarization measured by
the FPP are given as follows,
\begin{subequations}
\label{eq:simul}
\begin{eqnarray}
{p'}_{N''}&=&{p'}_{N'}\nonumber\\
&=&\frac{1}{1+p_NA_N}(P+D_{N'N}p_N), \label{eq:simula}\\
{p'}_{S''}&=&{p'}_{S'}\cos\chi+{p'}_{L'}\sin\chi \nonumber\\
&=&\frac{1}{1+p_NA_N}
[(D_{S'S}p_S+D_{S'L}p_L)\cos\chi \nonumber\\
&&+(D_{L'S}p_S+D_{L'L}p_L)\sin\chi].\label{eq:simulb}
\end{eqnarray}
\end{subequations}

In the measurement at 0$^\circ$, the projectile helicity
frame and the outgoing particle helicity frame are
identical. In this case, off diagonal components of the
PT-observables, analyzing power, and vector polarization vanish 
($D_{SL}=D_{LS}=A_N=P=0$) due to the spatial rotational symmetry,
and Eq.~(\ref{eq:simul}) reduces to
\begin{subequations}
\label{eq:sim0deg}
\begin{eqnarray}
{p'}_{N''}&=&D_{N'N}p_N=D_{NN}p_N, \label{eq:sim0a}\\
{p'}_{S''}&=&D_{S'S}p_S\cos\chi+D_{L'L}p_L\sin\chi \nonumber\\
&=&D_{SS}p_S\cos\chi+D_{LL}p_L\sin\chi.
\label{eq:sim0b}
\end{eqnarray}
\end{subequations}
The PT-observables with non-primed suffixes ($D_{IJ}$) in
Eq.~(\ref{eq:sim0deg}) are defined in the projectile helicity
frame. The $D_{IJ}$'s are related to PT-observables with primed suffixes
($D_{I'J}$) defined in the projectile helicity frame and the outgoing
particle helicity frame by a spatial rotation of the coordinate system
of outgoing particles. The $D_{IJ}$'s naturally become identical with
the $D_{I'J}$'s at 0$^\circ$. 

Although the transverse diagonal components of
PT-observables have a simple relation $D_{SS}=D_{NN}$ at 0$^\circ$,
the $D_{SS}$ and $D_{NN}$ are treated as independent observables in
this experiment. The reason is that the acceptance of the spectrometer
Grand Raiden is not symmetrical with respect to vertical and
horizontal directions ($|\theta_x|\le20$ mrad, $|\theta_y|\le35$ mrad) and
this difference breaks the relation $D_{SS}=D_{NN}$ by $\approx$10\%
at most.

The measurements were repeated using vertically polarized
($p_S=p_L=0$) and horizontally polarized ($p_N=0$) beams
independently. In the measurements with a vertically polarized
beam, the analyzing power was deduced from the asymmetry of the cross
section by reversing the beam polarization. The vector
polarization $P$ and $D_{N'N}$ were deduced from simultaneous
equations about $p'_{N''}$, which were obtained from
Eq.~(\ref{eq:simula}) for each polarizing direction in the reversing
process. 

\begin{figure}
\includegraphics[scale=0.85]{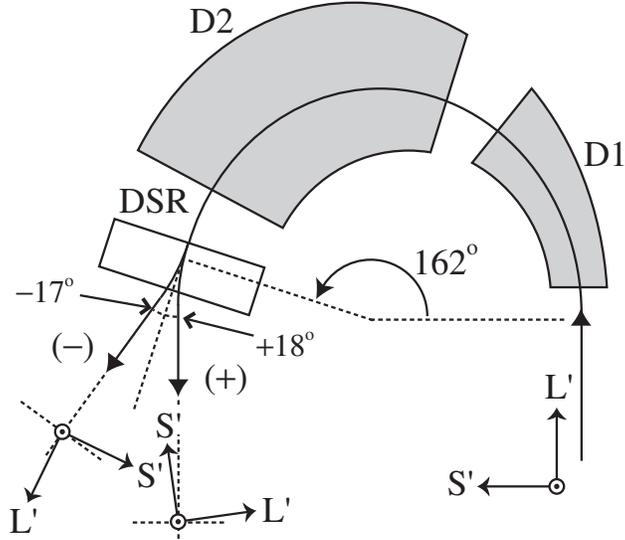}
\caption{\label{DSR}
Schematic description of the spin precession of horizontally polarized 
protons in the spectrometer Grand Raiden. Quadrupole and sextupole
magnets of Grand Raiden are not shown for simplicity. For details
see text.}
\end{figure}

Following Eqs.~(\ref{eq:simulb}) and (\ref{eq:sim0b}), two and four
independent measurements of $p'_{S''}$ with horizontally polarized
beams are generally required to obtain all horizontal
PT-observables at 0$^\circ$ and at other finite angles, respectively.
On the basis of the mathematical considerations mentioned above, we
measured $p'_{S''}$ under the two independent conditions with beam
polarizations perpendicular to each other in the horizontal plane. In
addition, for the measurements at
finite angles, we used a special dipole magnet for
spin rotation (DSR) \cite{MAMO89} to increase the number of
independent measurements. The DSR is a magnet, which bends protons by
$+18^\circ$ or $-17^\circ$ just in front of the focal plane of the
spectrometer Grand Raiden as shown in Fig.~\ref{DSR}. The bending
angles of scattered protons of the central ray are $180^\circ$ and
$145^\circ$ for positive and negative polarities of the DSR,
respectively. Then, the spin precession angles of 392 MeV protons in
Grand Raiden are $\chi^{(+)}=458^\circ$ and
$\chi^{(-)}=369^\circ$. The four independent measurements at finite
angles were achieved by measuring the $p'_{S''}$ with both beam
polarization axes for each DSR polarity. In the 0$^\circ$ measurement,
the DSR was also used as a steering magnet with a bending angle of
$1^\circ$-$2^\circ$ in order to correctly guide the proton beam into
the beam dump. In this case, the spin precession angle in the
spectrometer is $\chi\approx412^\circ$ determined by the normal
bending angle of $162^\circ$ of Grand Raiden.

The reliability of our measurements was checked by
measuring PT-observables of proton-proton elastic scattering.
We simultaneously measured protons scattered from hydrogen and
oxygen in the ice target at $\theta_{lab}=6^\circ$-$12^\circ$ since
the protons are still within the momentum acceptance of Grand
Raiden. The measured PT-observables for proton-proton elastic
scattering agreed well with the result of the SAID calculation.

\section{Theoretical Consideration}

Microscopic distorted-wave impulse approximation (DWIA) calculations
for $(p,p')$ reactions were performed using the computer codes
DWBA98 and DWBB98 \cite{RYNA98}. The effective nucleon-nucleon
interaction derived by Franey and Love (FL) \cite{FRAN85} at 425 MeV
was used in the calculations. The global Dirac optical-model
potential was used in the Schr\"{o}dinger equivalent form \cite{HAMA90}. 
This potential gives a good description for existing experimental
data of elastic scattering on $^{16}$O at 400 MeV \cite{CHAN85}. The
one-body transition density from the shell model calculation by Brown
\cite{BROW01} was used in the present work. This shell model
calculation was performed within
the (0+2) $\hbar\omega$ and (1+3) $\hbar\omega$ configuration spaces
for positive and negative parity states, respectively, with an 
interaction based on the WBP \cite{WARB92} and CD Bonn potential
\cite{MACH96}. The
single-particle radial wave functions were obtained for a
harmonic-oscillator potential with a size parameter of
$\alpha=0.588$ fm$^{-1}$. The calculated observables were averaged
over the acceptance of Grand Raiden ($|\theta_x|\le20$ mrad,
$|\theta_y|\le35$ mrad) weighted by the calculated cross sections for
comparison with the experimental data.

The spin-flip cross section $(d\sigma/d\Omega({\Delta}S=1))$ and 
non-spin-flip cross section $(d\sigma/d\Omega({\Delta}S=0))$ can be
defined by 
\begin{widetext}
\begin{subequations}
\label{eq:sigma}
\begin{eqnarray}
\frac{d\sigma}{d\Omega}({\Delta}S=1)
&=&\frac{3-(D_{SS}+D_{NN}+D_{LL})}{4}
\left(\frac{d\sigma}{d\Omega}\right) 
\equiv\Sigma\cdot\left(\frac{d\sigma}{d\Omega}\right),\label{eq:sigmasf}\\
\frac{d\sigma}{d\Omega}({\Delta}S=0)
&=&\frac{1+(D_{SS}+D_{NN}+D_{LL})}{4}
\left(\frac{d\sigma}{d\Omega}\right) 
\equiv(1-\Sigma)\cdot\left(\frac{d\sigma}{d\Omega}\right),\label{eq:sigmans}\\
\nonumber
\end{eqnarray}
\end{subequations}
\end{widetext}
where $d\sigma/d\Omega$ is a differential cross section. 
PT-observables in Eq.~(\ref{eq:sigma}) are defined in the projectile
helicity frame. It is known that the $\Sigma$ value in
Eq.~(\ref{eq:sigma}) is unity for spin-flip and zero for non-spin-flip
transitions at forward scattering angles where the spin-orbit
interaction is negligible \cite{SAKA98,SUZU00b}. This rule is well
established for unnatural isovector transitions. For natural parity
transitions, it is valid within 5\% accuracy for
$\theta\le5^\circ$. 

To verify the applicability of this rule, $d\sigma/d\Omega$ and
$(1-\Sigma){\cdot}d\sigma/d\Omega$ were calculated by DWIA for
isovector 1$^-$ states generated in the shell model space and were
compared with the calculated B(E1) values, which are good measures for
non-spin-flip transition strengths. The IVGDR is strongly excited by
the Coulomb interaction at 0$^\circ$. The $d\sigma/d\Omega$ in 
Coulomb excitation decreases with increasing excitation energy since
the virtual photon flux during the collision becomes rapidly small as
a function of energy. Therefore, all the calculations were performed
at an excitation energy of $E_x=15$ MeV in order to fix the kinematic
conditions. 

\begin{figure}
\includegraphics[scale=0.8]{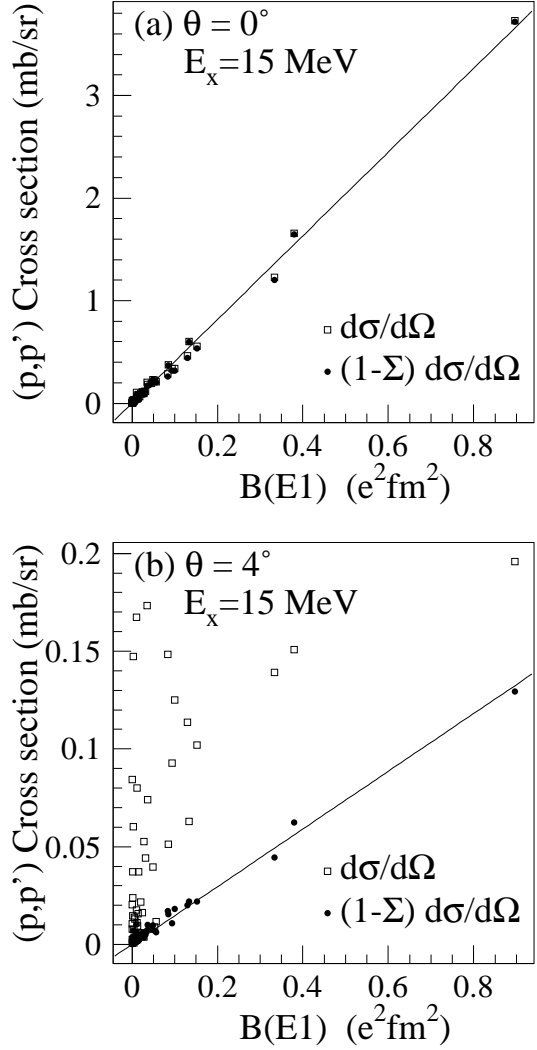}
\caption{\label{be1cs}
Two-dimensional scatter plots of cross sections ($d\sigma/d\Omega$,
$(1-\Sigma){\cdot}d\sigma/d\Omega$) at 0$^\circ$ (a) and 4$^\circ$ (b)
calculated for 1$^-$ shell model states. 
The open squares and the solid circles indicate cross sections
($d\sigma/d\Omega$) and non-spin-flip cross sections
($(1-\Sigma){\cdot}d\sigma/d\Omega$), respectively. The solid lines
(shown to guide the eye) indicate proportional relations between B(E1)
and $(1-\Sigma){\cdot}d\sigma/d\Omega$.
}
\end{figure}

The results are shown in the scatter plots of Fig.~\ref{be1cs}. 
The non-spin-flip strengths are dominant compared to the spin-flip
strengths at 0$^\circ$ due to the Coulomb excitation of the
IVGDR. Therefore, the strong linear correlation in the
scatter plots at $\theta=0^\circ$ (shown in
Fig.~\ref{be1cs}(a)) is interpreted as indication that the cross
sections observed at $0^\circ$ are nearly proportional to the E1
transition strengths. The non-spin-flip cross sections are quenched at
backward angles due to the destructive interference effect between the
Coulomb ($V_c$) and isovector ($V_\tau$) interactions. Non-spin-flip
cross sections have values much smaller than the cross sections at
$\theta=4^\circ$ as shown in Fig.~\ref{be1cs}(b). However, the
correlation between  non-spin-flip cross sections and B(E1)'s is still
linear. Thus, we conclude that the transition strengths are reasonably
separated into the spin-flip and non-spin-flip components by using
$\Sigma$ even at 4$^\circ$.

\section{Results and Discussion}

\begin{figure}
\includegraphics[scale=0.6]{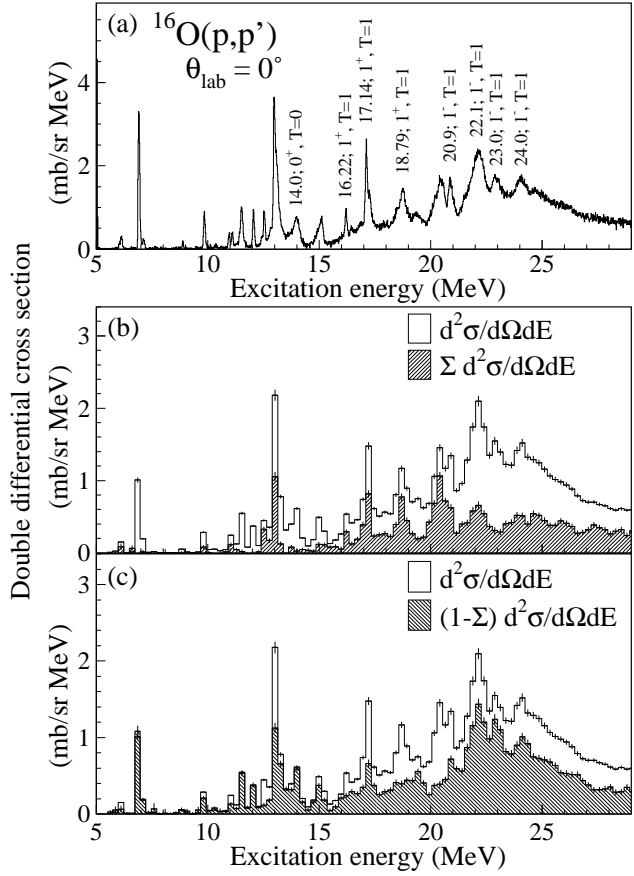}
\caption{\label{pt000}
Double differential cross sections for the $^{16}$O$(p,p')$ reaction
at $E_{p}=392$ MeV and $0^\circ$. (a) $^{16}$O$(p,p')$
spectrum $d^2\sigma/d{\Omega}dE$. (b) Spin-flip component
${\Sigma}{\cdot}d^2\sigma/d{\Omega}dE$ is compared with
$d^2\sigma/d{\Omega}dE$. (c) Non-spin-flip component
$(1-\Sigma){\cdot}d^2\sigma/d{\Omega}dE$ is compared with
$d^2\sigma/d{\Omega}dE$.}
\end{figure}

\begin{figure}
\includegraphics[scale=0.6]{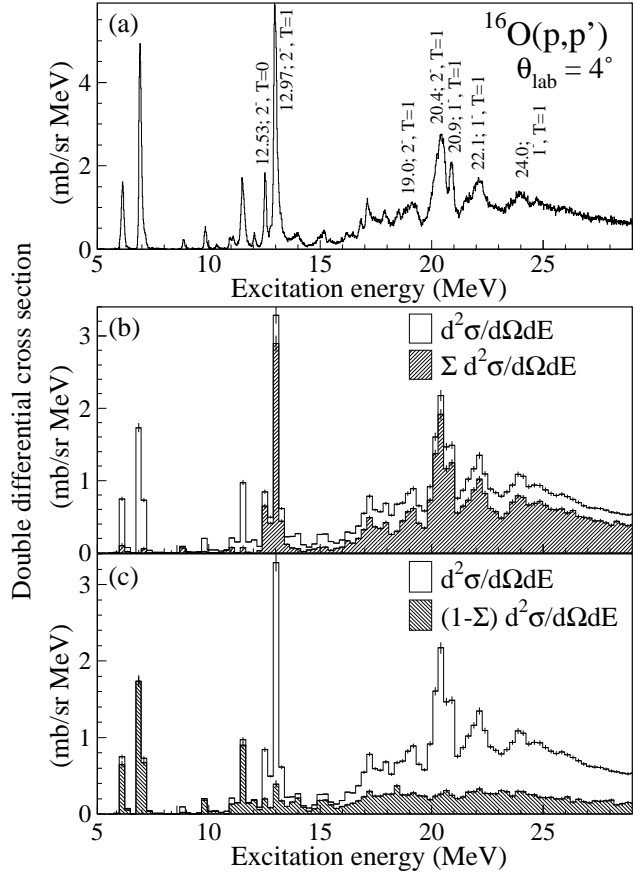}
\caption{\label{pt040}
Same as Fig.~\ref{pt000}, but at a laboratory angle of 4$^\circ$.
The bumps at $E_x=$ 19.0, 20.4, 20.9, 22.1, and 24.0 MeV are
identified to be due to ${\Delta}L=1$ transitions. Note that the bump
at 23.0 MeV seen in Fig.~\ref{pt000} is missing.
}
\end{figure}

\begin{figure}
\includegraphics[scale=0.6]{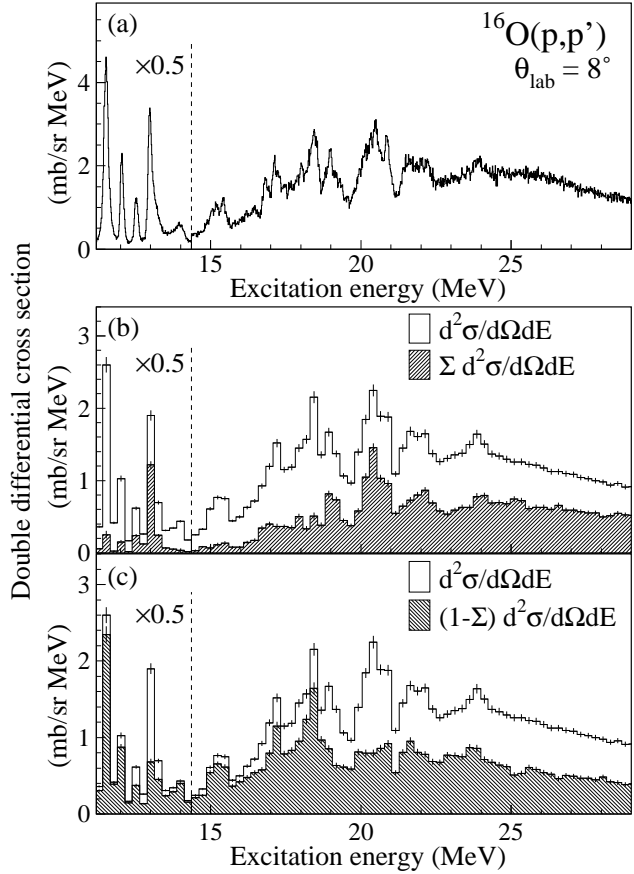}
\caption{\label{pt080}
Same as Fig.~\ref{pt000}, but at a laboratory angle of 8$^\circ$.
The spectra below $E_x=14.4$ MeV are scaled down by a factor of 0.5. 
}
\end{figure}

The double differential cross sections at $\theta_{lab}=0^\circ$,
4$^\circ$, and 8$^\circ$ for the $^{16}$O$(p,p')$ reaction at
$E_{p}=392$ MeV are shown in Figs.~\ref{pt000}, \ref{pt040}, and
\ref{pt080}, respectively. At 8$^\circ$, the $^{16}$O$(p,p')$ spectra
are obscured in the energy region of $E_x=$6$-$11 MeV due to the large
background originating from hydrogen in the ice target. Therefore, the
spectra in Fig.~\ref{pt080} are only shown for the energy region of
$E_x=$11.2$-$29 MeV.

All low-lying discrete peaks observed between 6.05 MeV and
13.09 MeV have been identified as those of known transitions
\cite{TILL93}. Table~\ref{tab:disc} lists the $0^\circ$
cross sections in the center of mass system for these known discrete
states. In the measurement where the central ray is set at $0^\circ$,
the average angle of the acceptance of the spectrometer is
1.2$^\circ$. The cross sections were
obtained by fitting the $^{16}$O$(p,p')$ spectrum at $0^\circ$. In
the fitting procedure, Lorentzian functions with central energies and
widths taken from Ref.~\cite{TILL93} were used. The Lorentzian
functions were folded by using a peak shape taken from the
narrow states at $E_x=$ 6.92 and 7.12 MeV. Although broad resonance
states at $E_x=$ 9.59, 11.26, and 11.60 MeV were taken into account to
improve the fit, cross sections of the transitions to these states
are not shown in Table~\ref{tab:disc} because of the large
uncertainties in the fit. Since the peaks of the broad states
are relatively small, the inclusion of the broad states into the fit
gives no significant influence in estimating the peak area of the
extracted states. The 11.10 MeV state was not separated from the
neighboring state at 11.08 MeV. Similarly, the 13.09 MeV state was not
separated from the 13.02 MeV state. Therefore, the summed values of
the cross sections for these neighboring states are shown in
Table~\ref{tab:disc}. 

\begin{table}
\caption{\label{tab:disc}Discrete levels observed in the
$^{16}$O$(p,p')$ reaction at $E_{p}=392$ MeV. Excitation energies
($E_x$), spin-parities ($J^\pi$), isospins ($T$), widths ($\Gamma$),
and level half-lives ($\tau_{1/2}$) are taken from
Ref.~\cite{TILL93}. Cross sections in the center of mass system were
obtained by fitting the $^{16}$O$(p,p')$ spectrum at $0^\circ$ (see
text). }
\begin{ruledtabular}
\begin{tabular}{cccc}
$E_x$ (MeV) &  $J^\pi$; $T$ & $\Gamma$ or $\tau_{1/2}$ & 
\begin{tabular}{c}
$d\sigma/d\Omega(0^\circ)_{c.m.}$\\
(${\rm \mu}$b/sr)
\end{tabular}
\\
\hline
 6.05 & $0^+$; 0 & $\tau_{1/2}=$ 67   $\pm$ 5    ps &   10 $\pm$ 1 \\
 6.13 & $3^-$; 0 & $\tau_{1/2}=$ 18.4 $\pm$ 0.5  ps &   23 $\pm$ 1 \\
 6.92 & $2^+$; 0 & $\tau_{1/2}=$ 4.7  $\pm$ 0.13 fs &  249 $\pm$ 7 \\
 7.12 & $1^-$; 0 & $\tau_{1/2}=$ 8.3  $\pm$ 0.5  fs &   19 $\pm$ 1 \\
 8.87 & $2^-$; 0 & $\tau_{1/2}=$ 125  $\pm$ 11   fs &   12 $\pm$ 1 \\
 9.59 & $1^-$; 0 & $\Gamma=$ 420  $\pm$ 20   keV    &    $-$       \\
 9.85 & $2^+$; 0 & $\Gamma=$ 0.62 $\pm$ 0.1  keV    &   64 $\pm$ 2 \\
10.36 & $4^+$; 0 & $\Gamma=$ 26   $\pm$ 3    keV    &   11 $\pm$ 1 \\
10.96 & $0^-$; 0 & $\tau_{1/2}=$ 5.5  $\pm$ 3.5  fs &   22 $\pm$ 2 \\
\begin{tabular}{c}
11.08 \\
11.10
\end{tabular}&
\begin{tabular}{c}
$3^+$; 0 \\
$4^+$; 0
\end{tabular}&
\begin{tabular}{c}
$\Gamma<$ 12 keV             \\
$\Gamma=$ 0.28 $\pm$ 0.05 keV
\end{tabular}& 
$\Biggr\}$\hspace{5pt} 24 $\pm$ 2\hspace{15pt}\mbox{ }\\
11.26 & $0^+$; 0 & $\Gamma=$ 2.5 MeV                &     $-$     \\
11.52 & $2^+$; 0 & $\Gamma=$ 71   $\pm$ 3    keV    &  137 $\pm$ 5 \\
11.60 & $3^-$; 0 & $\Gamma=$ 800  $\pm$ 100  keV    &     $-$      \\
12.05 & $0^+$; 0 & $\Gamma=$ 1.5  $\pm$ 0.5  keV    &   72 $\pm$ 3 \\
12.44 & $1^-$; 0 & $\Gamma=$ 91   $\pm$ 6    keV    &   36 $\pm$ 2 \\
12.53 & $2^-$; 0 & $\Gamma=$ 111  $\pm$ 1     eV    &   57 $\pm$ 2 \\
12.80 & $0^-$; 1 & $\Gamma=$  40  $\pm$ 4    keV    &   24 $\pm$ 2 \\
12.97 & $2^-$; 1 & $\Gamma=$ 1.34 $\pm$ 0.04 keV    &  176 $\pm$ 6 \\
\begin{tabular}{c}
13.02 \\
13.09
\end{tabular}&
\begin{tabular}{c}
$2^+$; 0 \\
$1^-$; 1 
\end{tabular}&
\begin{tabular}{c}
$\Gamma=$ 150  $\pm$ 10   keV \\
$\Gamma=$ 130  $\pm$ 5    keV
\end{tabular}&
$\Biggr\}$\hspace{5pt}{498 $\pm$ 14}\hspace{3pt}\mbox{ }
\end{tabular}
\end{ruledtabular}
\end{table}

It is known that low-lying states are mostly non-spin-flip states,
called natural parity states. Therefore, these states are only 
observed in our non-spin-flip spectrum
($(1-\Sigma){\cdot}d^2\sigma/d{\Omega}dE$). 
Excitations of the isovector states in $^{16}$O begin with $E_x=12.8$
MeV. The spin-flip states observed at 0$^\circ$ above the threshold
energy ($E_x=12.8$ MeV) are expected to be mainly due to the isovector
transitions, since the isovector spin term ($V_{\sigma\tau}$) in the
effective interaction is much stronger than the isoscalar spin term
($V_{\sigma}$) at small momentum transfer. In fact, the spin-flip
spectrum at $0^\circ$ shown in Fig.~\ref{pt000}(b) is quite similar to
the high resolution $^{16}$O$(p,n)$ spectra near 0$^\circ$ reported in
Refs.~\cite{FAZE82,WATS94} if the threshold energy for the isovector
transitions is taken into account.

A state at $E_x\approx13$ MeV is strongly excited with both 
spin-flip and non-spin-flip components at 0$^\circ$. The spin-flip
component of this state significantly increases at 4$^\circ$ as
shown in Fig.~\ref{pt040}(b). We conclude that this state consists of
the mixing of three discrete states; an isovector 1$^-$ state at 13.09
MeV which is strongly excited by the Coulomb interaction at 0$^\circ$,
an isoscalar 2$^+$ state at 13.02 MeV, and an isovector 2$^-$ state at
12.97 MeV. Three $1^+$ states at 16.22, 17.14, and 18.79 MeV 
observed in electron scattering \cite{KUCH83} are also seen in the
spin-flip spectra. The 14 MeV state, which has been tentatively
assigned as 1$^+$ or 0$^+$ in Ref.~\cite{DJAL87}, has no spin-flip
strength at 0$^\circ$. Thus, this state is inferred to be a 0$^+$ or
1$^-$ natural parity state. Most probably, this state could correspond
to the 0$^+$ state reported in the electron scattering experiment by
Hyde-Wright \cite{HYDE84}.

\subsection{Non-spin-flip transitions}
\begin{figure}
\includegraphics[scale=0.6]{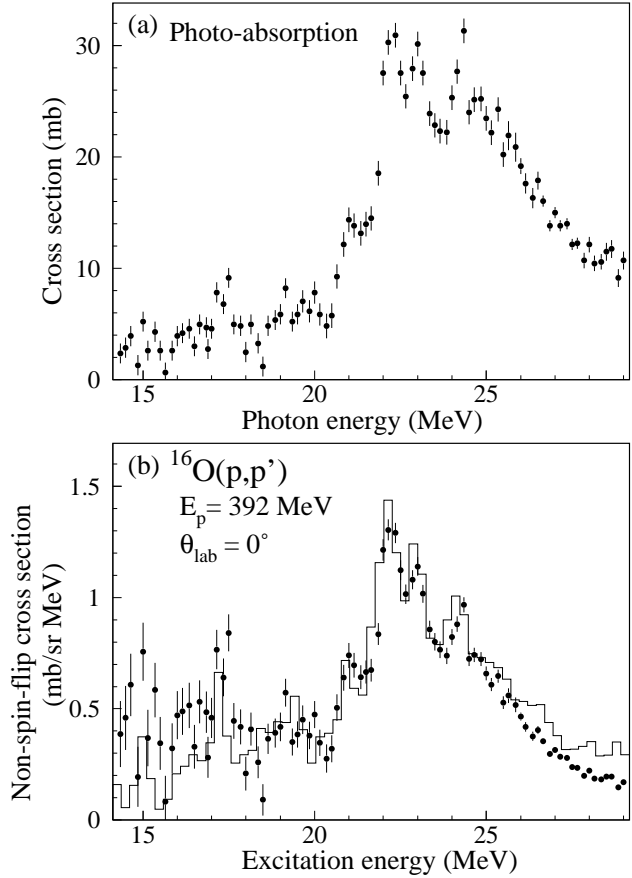}
\caption{\label{gamma}
(a) The photo-absorption spectrum from Ref.~\cite{AHRE75}.
(b) Comparison of the non-spin-flip $^{16}$O$(p,p')$ spectrum at
0$^\circ$ (solid lines) and the corresponding converted
photo-absorption spectrum (solid circles). See text for details.
Measurement errors for the non-spin-flip spectrum are not shown in
this figure for simplicity.
}
\end{figure}

Four broad resonance states at 20.9, 22.1, 23.0, and 24.0 MeV are
observed in the non-spin-flip spectrum at 0$^\circ$. The cross
sections of these non-spin-flip states are forward peaked which
could be characterized as IVGDR or ${\Delta}L=0$ transitions. The
IVGDR in $^{16}$O has been already well studied with electromagnetic
probes. The total photo-absorption cross sections from
Ref.~\cite{AHRE75} are shown in Fig.~\ref{gamma}(a) as a function of
photon energy. The relation between
photo-absorption cross sections and the reduced E1 transition matrix
element B(E1) is given by Ref.~\cite{BOHR75},
\begin{equation}
\int \sigma_{abs}dE=\frac{16\pi^3}{9{\hbar}c}S(E1)
=\frac{16\pi^3}{9{\hbar}c}\sum_{a}(E_a-E_0)B(E1;0{\to}a),
\label{eq:bohr}
\end{equation}
where S(E1) is the first energy moment of the B(E1) value.
By using Eq.~(\ref{eq:bohr}) and assuming the linear proportionality
between the B(E1) value and non-spin-flip cross sections at 0$^\circ$,
our results can be directly compared to those from the photo-absorption
experiment. The conversion factors from B(E1) to non-spin-flip cross
sections were deduced from DWIA calculations at each excitation energy
to correct kinematic effects. Non-spin-flip 1$^-$ transitions were
assumed in the calculation. Their wave functions were obtained from a
normal-mode procedure using the computer code NORMOD
\cite{WERF98}. After multiplying the converted photon absorption
spectrum by a factor of 1.3, the converted spectrum agrees well with
the non-spin-flip spectrum from the $(p,p')$ experiment as seen in
Fig.~\ref{gamma}(b). Contributions from the quasi-free process and the
monopole resonance have to be considered, since they might reduce the
difference of the excitation strengths between the $(p,p')$ results
and the calculation which includes only the E1 transitions. The
continuum due to the quasi-free scattering and the isoscalar giant
monopole resonance (ISGMR; $\Delta T=0$, $\Delta S=0$, $\Delta L=0$)
are expected at the same energy region as the IVGDR in the $(p,p')$
spectrum. However, the bumps due to the IVGDR are still observed in
the residual spectrum after the calculated cross sections are
subtracted from the non-spin-flip cross sections without any
normalization factor. Thus, the normalization factor of 1.3 is
required to explain the shape of the spectra. It is not clear why this
normalization factor is needed.

\begin{figure}
\includegraphics[scale=0.6]{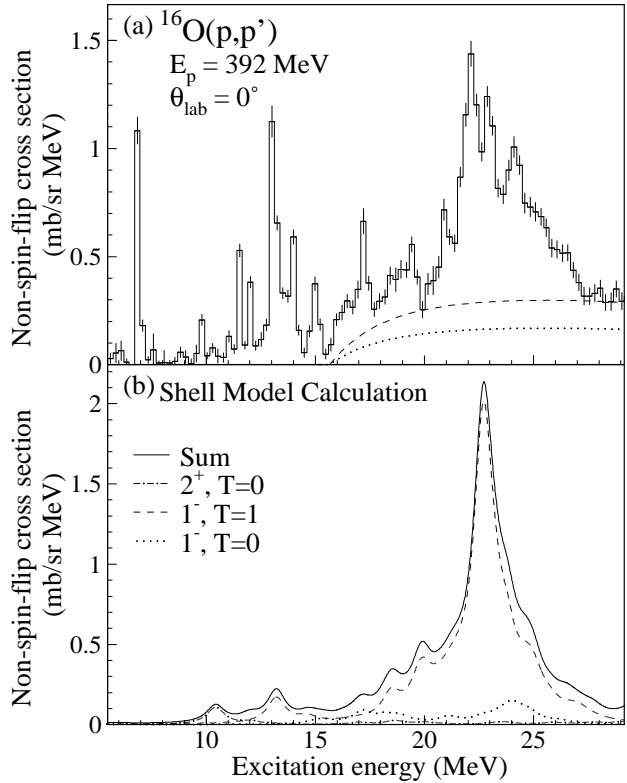}
\caption{\label{sm000ns}
(a) Measured non-spin-flip spectrum in the $^{16}$O$(p,p')$ reaction at
0$^\circ$. The dashed and dotted lines show an empirical estimation of
the quasi-free continuum (see text). (b) Calculated non-spin-flip
spectra obtained from the shell model calculation \cite{BROW01}.
The solid line shows the sum of all the transitions up to
${\Delta}J=4$, although the contributions from ${\Delta}J=0$, 3, and 4
are small. 
}
\end{figure}

We performed DWIA calculations using the shell model wave functions of
Ref.~\cite{BROW01} and the FL interaction of Ref.~\cite{FRAN85}. The
calculated non-spin-flip cross sections were folded by assuming that
each shell model state had a Lorentzian shape with a width of 1.0
MeV. In Fig.~\ref{sm000ns}, the measured spectrum for the
non-spin-flip cross sections at 0$^\circ$ is compared with those
estimated from the DWIA calculation. The gross structure of the
calculated spectrum was found to be similar to the observed one. The
excitation energies of the discrete isovector $1^-$ state at
$E_x=13.09$ MeV and the IVGDR located in the region of
$E_x=$20.9$-$24 MeV are well explained. The IVGDR exhausts most of
the non-spin-flip transition strengths in the calculation, while
contributions from ${\Delta}L=0$ transitions discussed above are
small. Recently, Lui {\it et al.} identified significant isoscalar
E0 strengths exhausting $48\pm10\%$ of the energy-weighted sum
rule (EWSR) in the region of $E_x=$11$-$40 MeV by using the
$^{16}$O($\alpha$, $\alpha$') reaction \cite{LUI01}. The shell model
calculation by Brown predicts isoscalar 0$^+$ strengths with $39.6\%$
of the EWSR in the same energy region. Nevertheless, the summed
$(p,p')$ cross section for the 0$^+$ states is less than $7\%$ of that
for the IVGDR at 0$^\circ$ in our calculation. One of the reasons why the
isoscalar 0$^+$ excitations are weak is that the isoscalar spin
independent term ($V_0$) in the effective interaction becomes small
around $E_p=400$ MeV.

The summed value of the calculated non-spin-flip cross
sections up to $E_x=29$ MeV is 7.57 mb/sr in the laboratory frame,
while the experimental value is 9.47 $\pm$ 0.04 mb/sr. It should be
noted that the experimental value includes a contribution from the
quasi-free process. If we tentatively assume that a smooth continuum
due to the quasi-free process begins with the neutron emission
threshold and increases with excitation energy as shown by a dashed
line in Fig.~\ref{sm000ns}(a), the non-spin-flip cross section of the
quasi-free process is estimated to be 3.35 mb/sr and the rest of the
non-spin flip cross section due to the resonant processes is 6.12
mb/sr, which is smaller by 20 $\%$ than the predicted value. In the
case that the magnitude of the quasi-free continuum is multiplied by a
factor of 0.6 as shown by the dotted line in Fig.~\ref{sm000ns}(a),
the experimental cross section for the resonant process becomes very
close to the theoretical one. 

\subsection{Spin-flip transitions}
In the spin-flip spectra shown in Figs.~\ref{pt000}(b) and
\ref{pt040}(b), broad bumps with ${\Delta}L=1$ were observed at $E_x=$
19.0, 20.4, 20.9, 22.1, and 24.0 MeV. Since the bumps at 19.0
and 20.4 MeV are not seen in the non-spin-flip spectra (see
Figs.~\ref{pt000}(c) and \ref{pt040}(c)), they are inferred to be 
excited by unnatural parity transitions, and correspond to the SDR
(2$^-$) reported in electron scattering \cite{GOLD70a,STRO70,KUCH83}. 
The other resonances at 20.9, 22.1, and 24.0 MeV, which are seen in
both the spin-flip and non-spin-flip spectra, could be due to $1^-$
excitations with a mixture of spin-flip and non-spin-flip
characters. The resonance at 20.9 MeV was assigned as 2$^-$ in
Ref.~\cite{DJAL87}, but our result favors the conclusion reported by
Ref.~\cite{STRO70} that the 20.9 MeV state is 1$^-$. A bump due to the
excitation of a $1^-$ resonance at $E_x=23.0$ MeV is clearly seen in
Fig.~\ref{pt000}(c). However, the corresponding bump is not observed
in the spin-flip spectrum at 4$^\circ$ (see
Fig.~\ref{pt040}(b)). Thus, we conclude that the $1^-$ transition to
the 23.0 MeV resonance is dominated by a non-spin-flip component.

\begin{figure}
\includegraphics[scale=0.6]{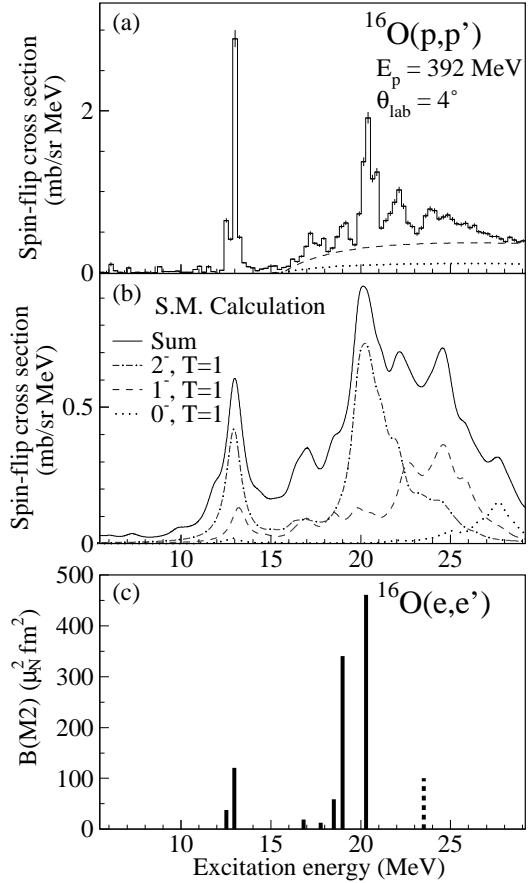}
\caption{\label{sm040sf}
(a) Measured spin-flip spectrum in the $^{16}$O$(p,p')$ reaction at
$\theta_{lab}=4^\circ$. The dashed and dotted lines show empirical
estimations of the quasi-free continuum (see text). (b) Calculated
spin-flip spectra
obtained from the shell model calculation \cite{BROW01}.
The solid line shows the sum of all the transitions up to
${\Delta}J=4$. (c) Measured M2 strength distribution in $^{16}$O
from Ref.~\cite{KUCH83}. The state at $E_x=23.5$ MeV (dashed line) was
tentatively assigned as $2^-$ in Refs.~\cite{GOLD70a,GOLD70b}.
}
\end{figure}

In Figs.~\ref{sm040sf}(a) and (c), the spin-flip spectrum at
$\theta_{lab}=4^\circ$ is compared with the results of electron
scattering experiments of Ref.~\cite{KUCH83}. The
spin-parity of the state at $E_x=23.5$ MeV was tentatively assigned as
$J^\pi=2^-$ in Refs.~\cite{GOLD70a,GOLD70b}. Our assignment for the
SDR (2$^-$) is consistent with the result from the electron scattering
experiments, but it is rather difficult to get a clear one-to-one
correspondence for the $2^-$ states at 16.82, 17.78, 18.50, and 23.5
MeV in the present experiment. The ratio of the strength of the 20.4
MeV and 19.0 MeV resonance is quite different from the result obtained
in electron scattering. This difference might be due to the
contribution of the orbital part in the electromagnetic interaction
which does not give a sizable effect in $(p,p')$ scattering at small
momentum transfer.

The spin-flip spectrum from the DWIA calculation described above is
presented in Fig.~\ref{sm040sf}(b). Similarly to the non-spin-flip
case, it is assumed that each shell model state has a Lorentzian shape 
with a width of 1.0 MeV. The calculation predicts a concentration of
discrete spin-flip strengths at $E_x\approx13$ MeV, which is
consistent with the experimental result. The theoretical and the
experimental cross sections at $\theta_{c.m.}=4.4^\circ$ 
(corresponding to $\theta_{lab}=4^\circ$) for the
discrete ${\Delta}L=1$ transitions are summarized in
Table~\ref{tab:sfdisc}. The two bumps observed at $E_x\approx13$ MeV
(see Fig.~\ref{sm040sf}(a)) are mainly due to the three ${\Delta}L=1$
states, {\it i.e.} the isoscalar 2$^-$ state at 12.53 MeV, the
isovector 2$^-$ at 12.97 MeV, and the isovector 1$^-$ at 13.09
MeV. The spin-flip cross sections for several weak states are not
deduced separately since the PT-observables for these states could not 
be extracted reliably. Their contributions are included in the values
for the neighboring strong states. The theoretical calculation
explains the cross sections ($d\sigma/d\Omega$) and the spin-flip
cross sections ($\Sigma{\cdot}d\sigma/d\Omega$) for the discrete
${\Delta}L=1$ states quite well with exception of the underestimated
strengths of the isovector 1$^-$ state at 13.09 MeV and the isoscalar
$1^-$ state at 12.44 MeV.

\begin{table*}
\caption{\label{tab:sfdisc}
Comparison of the measured cross sections
$({d\sigma}/{d\Omega})_{c.m.}$ and spin-flip cross sections
$(\Sigma\cdot{d\sigma}/{d\Omega})_{c.m.}$ with those from the DWIA
calculation based on the shell model wave functions \cite{BROW01}
for discrete ${\Delta}L=1$ transitions observed in the
$^{16}$O$(p,p')$ reaction at $\theta_{c.m.}=4.4^\circ$. Shell model
states weaker than 0.01 mb/sr are not shown. Spin-flip cross sections
for $E_x=$ 7.12 MeV and 10.96 MeV are not shown because of large
uncertainties of the PT-observables. 
}
\begin{ruledtabular}
\begin{tabular}{cccccccc}
\multicolumn{4}{c}{Experiment}&
\multicolumn{4}{c}{Theory}\\
$E_x$ & $J^\pi$; $T$ & $\frac{d\sigma}{d\Omega}(4.4^\circ)_{c.m.}$ &
$\Sigma\cdot\frac{d\sigma}{d\Omega}(4.4^\circ)_{c.m.}$ &
$E_x$ & $J^\pi$; $T$ & $\frac{d\sigma}{d\Omega}(4.4^\circ)_{c.m.}$ &
$\Sigma\cdot\frac{d\sigma}{d\Omega}(4.4^\circ)_{c.m.}$ \\
(MeV) &              & (${\rm \mu}$b/sr) &(${\rm \mu}$b/sr) &
(MeV) &              & (${\rm \mu}$b/sr) &(${\rm \mu}$b/sr) \\
\hline
7.12  & $1^-$; 0 & 48 $\pm$ 2 & $-$        & 6.04 & $1^-$; 0 & 41 & 18 \\
8.87  & $2^-$; 0 & 25 $\pm$ 1 & 19 $\pm$ 2 & 7.28 & $2^-$; 0 & 38 & 27 \\
10.96 & $0^-$; 0 & 25 $\pm$ 2 & $-$        & 9.84 & $0^-$; 0 & 32 & 32 \\
\begin{tabular}{c}
12.44 \\ 
12.53
\end{tabular}&
\begin{tabular}{c}
$1^-$; 0 \\
$2^-$; 0 
\end{tabular}&
\begin{tabular}{c}
37  $\pm$ 6\\
179 $\pm$ 30
\end{tabular}&
$\Biggr\}$\hspace{5pt}{164 $\pm$ 12} \hspace{10pt}\mbox{ }&
\begin{tabular}{c}
11.38\\
11.84
\end{tabular}&
\begin{tabular}{c}
$1^-$; 0 \\
$2^-$; 0 
\end{tabular}&
\begin{tabular}{c}
10 \\
189
\end{tabular}&
\begin{tabular}{c}
5 \\
131
\end{tabular}\\
\begin{tabular}{c}
12.80 \\ 
12.97
\end{tabular}&
\begin{tabular}{c}
$0^-$; 1 \\
$2^-$; 1 
\end{tabular}&
\begin{tabular}{c}
42  $\pm$ 2\\
534 $\pm$ 15
\end{tabular}&
$\Biggr\}$\hspace{5pt}{600 $\pm$ 24} \hspace{10pt}\mbox{ }&
\begin{tabular}{c}
12.90\\
12.94
\end{tabular}&
\begin{tabular}{c}
$0^-$; 1 \\
$2^-$; 1 
\end{tabular}&
\begin{tabular}{c}
26 \\
545
\end{tabular}&
\begin{tabular}{c}
26 \\
544
\end{tabular}\\
&&&&13.12&$2^-$; 0&39&25\\
\begin{tabular}{c}
13.02 \\ 
13.09
\end{tabular}&
\begin{tabular}{c}
$2^+$; 0 \\
$1^-$; 1 
\end{tabular}&
$\Biggr\}$\hspace{5pt}
490 $\pm$ 15 \hspace{10pt}\mbox{ }& 342 $\pm$ 20&
\begin{tabular}{c}
\\ 13.21
\end{tabular}&
\begin{tabular}{c}
\\ $1^-$; 1
\end{tabular}&
\begin{tabular}{c}
\\ 174
\end{tabular}&
\begin{tabular}{c}
\\ 167
\end{tabular}\\
\end{tabular}
\end{ruledtabular}
\end{table*}

In the region of giant resonances, the calculation reproduces the
experimental result that the 2$^-$ strength concentrates at an
excitation energy below the $1^-$ strength. This ${\Delta}J$ splitting
is expected due to the spin-orbit interaction, supporting the validity
of the present calculations. The strong resonance at 20.4 MeV is
predominantly due to a 2$^-$ transition, while the 22.1
MeV is due to both $2^-$ and $1^-$ transitions according to the
calculation. In addition, the shell model calculation
predicts a considerable 0$^-$ strength at higher excitation energies.
However, such a $0^-$ strength could not be separated reliably from
the quasi-free background. It is noteworthy to mention that a simple
1p-1h shell model calculation by Picklesimer and Walker \cite{PICK78}
has predicted the gross structure of the SDR similar to the recent
sophisticated calculation \cite{BROW01}, although the quenching
problem for spin excitations was not seriously discussed before
1980's.

The sum of the experimental spin-flip cross sections up to $E_x=29$
MeV, which includes the continuum, is
9.10 $\pm$ 0.03 mb/sr in the laboratory frame, while that of the
calculation is 7.57 mb/sr. Assuming a smooth quasi-free continuum as
shown in Fig.~\ref{sm040sf}(a) by the dashed line,
the spin-flip cross section due to the resonant processes is 5.06
mb/sr. This is smaller than the 7.57 mb/sr value from the
calculation. The best agreement between experimental and calculated
values is obtained if the estimated quasi-free continuum is
multiplied by a factor of 0.3 shown as dotted line in
Fig.~\ref{sm040sf}(a).

\section{Summary and Conclusion}
In the present $^{16}$O$(p,p')$ experiment, spin-flip and 
non-spin-flip transitions were separated by measuring the
polarization transfer (PT) observables. Strong peaks due to M1
transitions were observed at $E_x=16.22$, 17.14, and 18.77
MeV. The 14.0 MeV state, which was previously suggested to be excited
by an M1 transition, is found to have non-spin-flip nature. 

Non-spin-flip transitions with forward peaked cross sections were
observed at excitation energies between 20 MeV and 27 MeV. These
transitions are well reproduced by a calculation in
which the excitation strengths are converted from the
photo-absorption cross sections with a normalization factor of
1.3. Therefore, we conclude that the major part of the IVGDR
strength is exhausted in the resonance region of $E_x=$20$-$27
MeV. The shell model calculation by Brown \cite{BROW01} also supports
this conclusion. 
One may address a question that a significant strength due to the
$0^+$ excitation could, in principle, exist in the IVGDR
region. However, the contributions from the ISGMR are rather small
according to the calculation. 

Spin-flip strengths observed in the same energy region with the IVGDR
are found to be excited with ${\Delta}L=1$ angular momentum
transfer. The resonances observed at $E_x=$ 20.9, 22.1, and 24.0 MeV
carry both IVGDR and SDR ($1^-$) strengths. The resonances at
$E_x=19.0$ and 20.4 MeV are observed only in the spin-flip spectra and 
are therefore assigned to be $2^-$ states. The energies of strong
$2^-$ states observed in the present $(p,p')$ experiment agree well
with those of the $2^-$ states at $E_x=12.53$, 12.97, 19.0 and 20.3
MeV reported in electron scattering studies
\cite{KUCH83,GOLD70a,STRO70,GOLD70b}. However, the strength ratios of
the 19.0 and 20.4 MeV states are different. This may be attributed to
the intrinsic contribution of the orbital part in electron
scattering. We observed that the major part of the SDR ($2^-$)
strength in $^{16}$O is located at excitation energies below the SDR
($1^-$) as seen in the $A=12$ system \cite{GAAR84,YANG93,INOM98}.
This result is consistent with the expectation that the three spin
components of the SDR split in excitation energy in the order of
$E_x(2^-)<E_x(1^-)<E_x(0^-)$. The shell model calculation by
Brown \cite{BROW01} reproduces well the distribution of
the $2^-$ and $1^-$ spin-flip strengths in $^{16}$O measured in this
study. This calculation predicts the existence of the $0^-$ transition
at high excitation energies about 27 MeV. However, not enough evidence
to identify the 0$^-$ strength could be found in the experiment
because of its relatively weak excitation and the surrounding
quasi-free background. The experimental result from the present
$(p,p')$ study will be useful in estimating the neutrino absorption
cross sections by $^{16}$O in the supernova neutrino observatories.

\begin{acknowledgments}
The authors would like to thank Prof.~B.A.~Brown, Prof.~J.~Raynal,
Prof.~K.~Amos, Prof.~S.~Karataglidis, and Prof.~S.Y.~van~der~Werf for
valuable discussions of theoretical calculations. We gratefully
acknowledge the outstanding effort of the RCNP cyclotron staff for
providing the clean stable beam. This research program was supported
in part by the Research Fellowships of the Japan Society for the
Promotion of Science (JSPS) for Young Scientists, and by the
U.S.-Japan Cooperative Science Program of the JSPS.
\end{acknowledgments}

\end{document}